# Improving efficiency and stability for perovskite solar cell with diethylene glycol dimethacrylate modification

Xiaodan Wei [a,1], Shuyuan Zhang [a,1], Boyang Ma [a,1], Zehua Feng [a,1], chunhan Yu [a], Linyuan Peng [a], Zijian Hua [a], Jie Xu [a*]

[a] School of science, Xi'an University of Architecture and Technology, Xi'an, 710055, China

## Abstract

The humidity resistance is the key challenges that hinder the commercial application of perovskite solar cells (PSCs). Herein, we propose an ultra-thin acrylate polymer (diethylene glycol dimethacrylate, DGDMA) into perovskite films to investigate the influence of polymerized networks on stability. The monomer molecules containing acrylate and carbonyl groups were selected, and the effects of the polymerized were quantified with different concentration. The experimental results show that, when the concentration of DGDMA is 1 mg/ml, the PCE increases from 18.06% to 21.82%, which is optimum. The monomer molecules with carbonyl groups polymerize, they can chelate with uncoordinated $Pb^{2+}$ in perovskite films to improve the film quality, reduce the surface defect density to decrease non-radiative recombination, and also significantly enhance the humidity stability of PSCs.

**Keywords:**

Perovskite solar cells, Diethylene glycol dimethacrylate, Humidity resistance, Stability.

# 1. Introduction

Organic-inorganic lead halide perovskite solar cells (PSCs) has gradually become a newcomer in the field of photovoltaics, which has attracted the research of many researchers due to many advantages of the materials, such as low cost, simple production process and high optoelectronic properties[1-3]. In the last decade, the power conversion efficiency (PCE) of small-area PSC has broken from 3.8% to 27%[4,5]. However, due to the poor stability of perovskite film, the device has not reached the requirement that it can be used for commercialization.

In order to solve the above problems, additive engineering is considered as a feasible and effective technique to regulate crystal growth and passivate defects. Adding some polymers with hydroxyl, carbonyl and other groups, such as polyethylene glycol[6], polyvinyl pyridone[7], polyvinyl alcohol and polymethylmethacrylate[8], these polymers will form strong bonds with the noncoordinating ions, $Pb^{2+}$ or $I^-$, and will reduce the non-radiative recombination[9-11]. However, often these polymers interact too strongly with $PbI_2$ during the preparation process, which will form precipitation in the perovskite precursor unavoidably, which also reduces the grain size and makes the fabrication of homogeneous polymer perovskites very challenging. Therefore, the addition of small molecule monomers to the perovskite precursor was able to initiate cross-linking polymerization during thermal annealing[12-14]. Most of the reported high-efficiency PSCs fabrication processes are done in a nitrogen glove box, and the "one-step anti-solvent method" solution spin-coating is mainly utilized for the preparation of the devices, for which the dropping time of the anti-solvent and the dosage of the anti-solvent are all related to the perovskite precursor. Moreover, sometimes with the change of the atmosphere of the glove box, the process of anti-solvent needs to be adjusted precisely, so the spin-coating of anti-solvent is a crucial step in the process of preparation.

In this work, a functional monomer molecule, diethylene glycol dimethacrylate (DGDMA), is used to add to the anti-solvent chlorobenzene (CB) with an internal encapsulation strategy, which avoids the formation of precipitates with the perovskite precursor. In addition, the monomer molecules containing carbonyl groups in

DGDMA can chelate with the uncoordinated $Pb^{2+}$ within the perovskite film, which can try to avoid the problem of Pb leakage in perovskite, and can reduce the defect density, non-radiative recombination, and improve the performance of the device. The experimental results show that the morphology of perovskite film is the flattest when the concentration is 1 mg/ml, and the efficiency is also the highest which can reach 21.82%. This method can also improve the humidity stability of perovskite film through the water contact angle test. The study not only provides a high-efficiency optimization strategy, but also provides new idea for improving stability.

**2. Experimental**

*2.1 Materials and reagents*

Methylamium iodine (MAI, 99.99%), formamidinium iodide (FAI, 99.5%), cesium iodide (CsI, 99.99%), and 2PACZ were purchased from Xi'an Polymer Light Technology Corp (China). Lead diiodide ($PbI_2$, >99.99%) was purchased from TCI (Japan). Bathocuproine (BCP, 99.0%), and buckminsterfullerene ($C_{60}$, 99%) were purchased from Nichem (Taiwan). Besides, some reagents including chlorobenzene (CB, Extra Dry, 99.8%), DMF (anhydrous, 99.8%), and DMSO (anhydrous, 99.8+%) were purchased from commercial sources (Acros) and used as received. Silver was obtained from commercial sources (China New Metal Materials Technology Co. Ltd.) with high purity (99.99%)

*2.2 Device fabrication*

The specification of ITO conductive glass used in the experiment is 2.5 cm × 2.5 cm, and its conductive surface is used as the front side. The cleaning part was divided into four steps: ITO cleaning solution, ITO post-shabbling (deionized water) cleaning, ethanol-acetone mixture (1:1 volume) cleaning, and deionized water. The ITO cleaning solution cleaning and ethanol-acetone mixture cleaning were cleaned by ultrasonic waves for 15 minutes, and the remaining two steps were cleaned by ultrasonic waves for 10 minutes, respectively. After cleaning, in the ultra-clean table using infrared baking lamp irradiation until the ITO surface without water, the ozone treatment for 15 minutes after the rapid transfer to the glove box.

The prepared 2PACZ solution (dissolve 0.5 mg of 2PACZ into 1 ml of ethanol solution and stir for 3 hours at room temperature to obtain 0.5 mg/ml of hole transport layer solution) was filtered and spun-coated on the front side of the ITO glass with a homogenizer, 100 ul of hole transport layer solution was added dropwise to each glass slice with a rotational speed of 3000 rpm/s, and spun-coated for 30 s, in which the acceleration rate was 3000 rpm/s; and finally annealed, at 100 °C for 10 min.

1.5 M CsI solution preparation: 289.7 mg of CsI was dissolved into 1 ml of DMSO solution and stirred at 80 °C for 1 hour to obtain 1.5 M CsI solution. Preparation of perovskite precursor: 228.8 mg FAI, 668.3 mg $PbI_2$, 7.8 mg MABr, 28.0 mg $PbBr_2$, and 14.2 mg MACl were added into a mixed solvent consisting of 800 ul of DMF and 160 ul of DMSO, and stirred for about 30 min at room temperature until all the solutes were dissolved. A solution of perovskite precursor was obtained by adding 40 ul 1.5 M CsI solution and stirring for 3 h at room temperature. The prepared perovskite precursor was filtered, and the precursor solution was spin-coated on the hole transport layer 2PACZ film using a homogenizer, with 70 ul of perovskite precursor solution dropwise added to each glass slide, the rotational speed set at 4000 rpm/s, the spin-coating time at 40 s, where the acceleration was 800 rpm/s, and 200 ul of the counteracting solvent dropwise added to the substrate 7 s before the end of the spin-coating. 200 ul of counter-solvent was added onto the substrate 7 s before the end of spin-coating, and then the spin-coated substrate was quickly transferred to a heating table and annealed at 100 °C for 60 min. 1 mg of PEAI was dissolved in 1 ml of isopropanol (IPA) and stirred for about 3 hours at room temperature to obtain 1 mg/ml of PEAI solution. 2 mg of diethylene glycol dimethacrylate (DGDMA) was dissolved into 4 ml, 2 ml, 1 ml and 500 ul of chlorobenzene and stirred for about 3 hours at room temperature to obtain 0.5 mg/ml, 1 mg/ml, 2 mg/ml and 4 mg/ml solutions. The prepared PEAI solution was filtered and waited for the completion of perovskite annealing, and the substrate at the end of perovskite annealing was continued to be spin-coated by a homogenizer, with 100 ul of PEAI solution dropwise for each glass slide at a rotational speed of 3,000 rpm/s, and a spin-coating time of 30 s, where the acceleration was 3,000 rpm/s, and no annealing was required, and the subsequent device preparation operations were carried out directly. The annealed substrate was placed into a vacuum vaporizer, and the $C_{60}$ (voltage 3 V, film thickness of about 40 nm), BCP (voltage 2.2 V, film thickness of about 6 nm), and Ag (adjustable voltage, film thickness of about 90 nm) were successively thermally vaporized under an air pressure of $<4.5 \times 10^{-4}$ Pa to complete the preparation of perovskite solar cell devices.

*2.3 Measurement and characterization*

Field emission scanning electron microscope (SEM) including energy-dispersive X-ray (EDX) (Quanta 250, FEI, USA) was used to investigate the morphology. The crystalline structure on ITO substrate is studied by Xray diffractometer (XRD) (Bruker D8 ADVANCE) with Cu Kα radiation. The absorption was obtained by ultraviolet–visible spectrophotometer (HITACHI U-3010, Japan). The PV performance was estimated under an AAA solar simulator (XES-301S, SAN-EI), AM 1.5 G irradiation with an intensity of 100 mW/cm$^2$. The photocurrent density-voltage (*J-V*) curve was measured using Keithley (2602 Series Sourcemeter) with scan rate 0.01 V/s within the range from -0.1 to 1.2 V. The area of each device, calibrated by the shadow mask, was 9 mm$^2$. The PL lifetime and PL spectra were recorded in a FLS 920 Fluorescence Lifetime and Steady State Spectroscopy (Edinburgh Instruments, British).

## 3. Results and discussions

To investigate the impact of DGDMA modification on the morphology of perovskite films, this section first conducts SEM (Scanning Electron Microscope) testing and characterization (**Fig. 1**). Perovskite films were prepared by adding DGDMA at various concentrations (0 mg/ml, 0.5 mg/ml, 1 mg/ml, 2 mg/ml, 4 mg/ml) to the antisolvent CB, with the sample containing 0 mg/ml serving as the control group.

**Fig. 1a** shows that the surface of the control is uneven, with a few pores present. As the concentration of the DGDMA additive increases, the pores in the perovskite film disappear when the additive concentration reaches 0.5 mg/ml (**Fig. 1b**). When the additive concentration further increases to 1 mg/ml, the grain size distribution becomes more uniform, and the surface becomes even smoother (**Fig. 1c**). However, as the additive concentration continues to increase, **Fig. 1d and Fig. 1e**, when the concentration reaches 2 mg/ml and 4 mg/ml, the overall smoothness of the grains decreases, and the uniformity of grain size diminishes, with the grain size decreasing as well. Preliminary test results confirm that an appropriate additive concentration is more conducive to obtaining perovskite films with a more uniform grain size distribution.

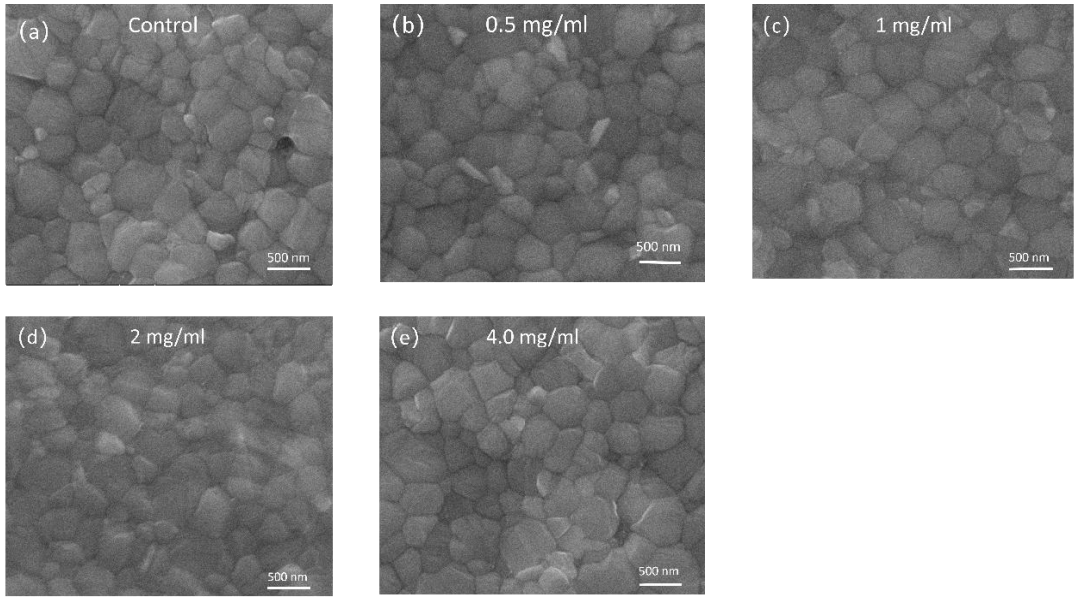

**Fig. 1.** SEM of perovskite films without and with DGDMA with different concentrations ((a) Control, (b) 0.5 mg/ml, (c) 1.0 mg/ml, (d) 2.0 mg/ml, (e) 4.0 mg/ml, respectivly).

XRD were conducted on perovskite films prepared with different concentrations of DGDMA (0 mg/ml, 0.5 mg/ml, 1.0 mg/ml, 2.0 mg/ml, 4.0 mg/ml), as shown in **Fig. 2**. All the perovskite films exhibited three high-intensity diffraction peaks at 14.1°, 28.4°, and 32.2°, corresponding to the (110), (220), and (310) crystal planes of the perovskite, respectively. The XRD results indicate that the addition of different concentrations of DGDMA did not alter the crystal structure of the perovskite. The full width at half maximum (FWHM) values of the perovskite films prepared with DGDMA concentrations of 0.5 mg/ml and 1 mg/ml decreased for the (110), (220), and (310) crystal planes. According to the Scherrer equation, a lower FWHM value corresponds to larger grain size. As the DGDMA concentration increased to 2.0 mg/ml and 4.0 mg/ml, the FWHM values increased, which is consistent with the changes observed in the SEM. Therefore, adding a small amount of DGDMA is beneficial for the crystallization of perovskite films and obtaining higher-quality perovskite films with a uniform distribution of grain sizes. However, excessive addition of DGDMA can degrade the film quality. Additionally, besides the three high-intensity diffraction peaks, there are also diffraction peaks at 12.8°, 19.7°, 24.3°, and 34.6°. The diffraction peak at 12.8° corresponds to the (001) crystal plane of $PbI_2$, and the presence of a small amount of $PbI_2$ can also have a passivating effect. The other diffraction peaks correspond to the (112), (211), and (312) crystal planes of the perovskite respectively. SEM and XRD show that when the concentration of added DGDMA is within 2.0 mg/ml, the perovskite films exhibit good uniformity in morphology and crystallinity. Therefore, for subsequent film testing, the concentration of added DGDMA should be controlled within 2.0 mg/ml.

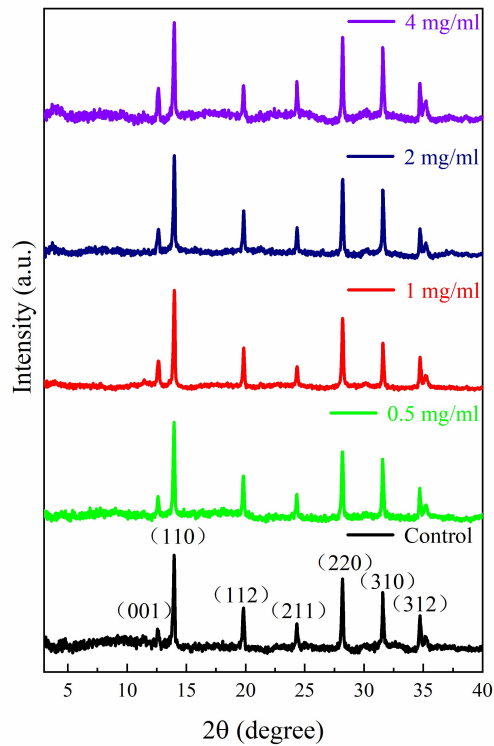

**Fig. 2.** XRD patterns of perovskite films before and after addition of DGDMA with different concentrations

To investigate the changes in optical properties of perovskite films before and after modification with DGDMA, PL spectroscopy was conducted on different perovskite films. These films were prepared by DGDMA at various concentrations (0 mg/ml, 0.5 mg/ml, 1 mg/ml, 2 mg/ml) to the antisolvent CB, as shown in **Fig. 3**. The sample with 0 mg/ml of DGDMA was designated as the control. The spectra indicate that the perovskite films modified with different concentrations of DGDMA all exhibit emission peaks near 790 nm, which aligns with the typical emission peak position of perovskites, as reported in the literature. When the concentration is 1 mg/ml, the emission peak reaches its highest intensity, suggesting that the introduction of DGDMA can reduce non-radiative recombination centers and effectively decrease the defects in the films. However, as the concentration increases, the intensity of the emission peak decreases significantly. This is due to the excess DGDMA remaining unreacted during the annealing process of the perovskite films, which then acts as non-radiative recombination, leading to a notable reduction in the PL intensity of the films.

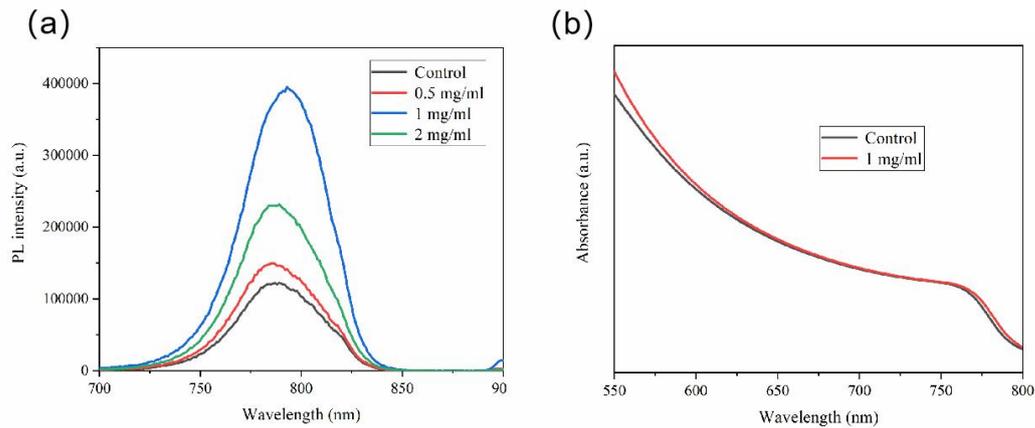

**Fig. 3.** (a) PL and (b) absorption spectra of perovskite films without and DGDMA with different concentrations.

Absorption spectrum were conducted on perovskite films prepared with different concentrations of DGDMA (0 mg/ml and 1 mg/ml), as shown in **Fig. 3b**. The absorption spectrum and absorption edge of the perovskite films with DGDMA are generally consistent with control, indicating that the introduction of DGDMA does not change the optical bandgap of the perovskite. The absorption intensity of the film slightly increases after adding DGDMA. The main reason for this is the enhanced compactness of the film, improved surface smoothness, and increased crystallinity upon the introduction of DGDMA. These improvements reduce the loss due to light reflection and enhance the light absorption characteristics of the films. This observation is also consistent with the SEM and XRD analysis results mentioned above.

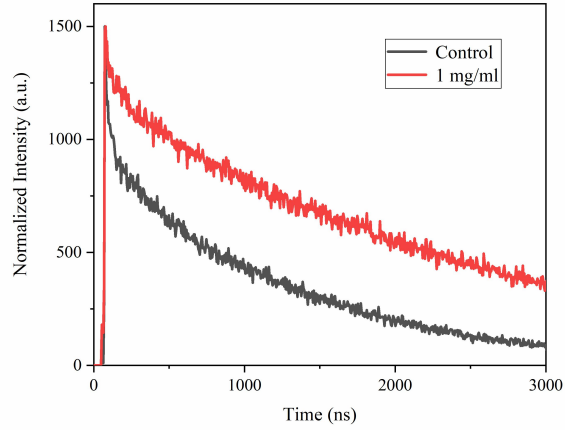

**Fig. 4.** TRPL of perovskite films without and with DGDMA.

Transient photoluminescence (TRPL) spectroscopy was conducted on perovskite films prepared with different concentrations of DGDMA (0 mg/ml and 1 mg/ml), as shown in **Fig. 4**. The TRPL decay curves were fitted using a biexponential decay model, as shown in Equation (1):

$$Y = A_1 \exp\left(-\frac{t}{\tau_1}\right) + A_2 \exp\left(-\frac{t}{\tau_2}\right) + y_0 \quad (1)$$

Where $A_1$ and $A_2$ are proportionality factors, $\tau_1$ represents the decay due to the recombination of photogenerated excitons at defects or grain boundaries, and $\tau_2$ represents the decay due to the exciton lifetime of the film itself. $y_0$ is a constant.

**Table 1.** Fitting parameters of time-resolved photoluminescence (TRPL) for perovskite films without and with DGDMA.

| Samples | $\tau_1$/ns | $\tau_2$/ns | $A_1$ | $A_2$ | $\tau_{average}$/ns |
|---|---|---|---|---|---|
| Control | 43.56 | 1295.05 | 0.02 | 0.98 | 1270.02 |
| 1 mg/ml | 62.56 | 2702.97 | 0.03 | 0.97 | 2623.75 |

After modification with DGDMA, $\tau_1$ increased from 43.56 ns to 62.56 ns, $\tau_2$ increased from 1295.05 ns to 2702.97 ns, and the average PL lifetime increased from 1270.02 ns to 2623.75 ns. The specific values are shown in **Table 1**. These results indicate that DGDMA modification can reduce the defect density of the film, inhibit non-radiative recombination processes, and extend the decay time constants.

This section explores the impact on the optoelectronic performance of PSCs prepared with different concentrations of DGDMA. The device structure employed is Glass/ITO/SAM（2PACZ）/Cs$_{0.05}$(FA$_{0.95}$MA$_{0.05}$)$_{0.95}$Pb(I$_{0.95}$Br$_{0.05}$)$_3$/C$_{60}$/BCP/Ag, which is a typical inverted planar heterojunction structure. SEM cross-sectional imaging of the PSCs is conducted, as shown in **Fig. 5a**. The image reveals that the perovskite films prepared after adding DGDMA penetrate the entire light-absorbing layer, which can reduce the defect density within the perovskite films. The close adhesion of the perovskite layer to both the upper and lower interfaces also enhances carrier extraction and transport capabilities, laying the foundation for subsequent improvements in the optoelectronic performance of the devices.

The key focus of research in perovskite solar cells is the photoelectric conversion efficiency of the devices. Therefore, *J-V* were conducted on devices prepared by adding different concentrations of DGDMA (0 mg/ml, 0.5 mg/ml, 1 mg/ml, 2 mg/ml) to investigate their impact on the photoelectric performance, as shown in **Fig. 5b**.

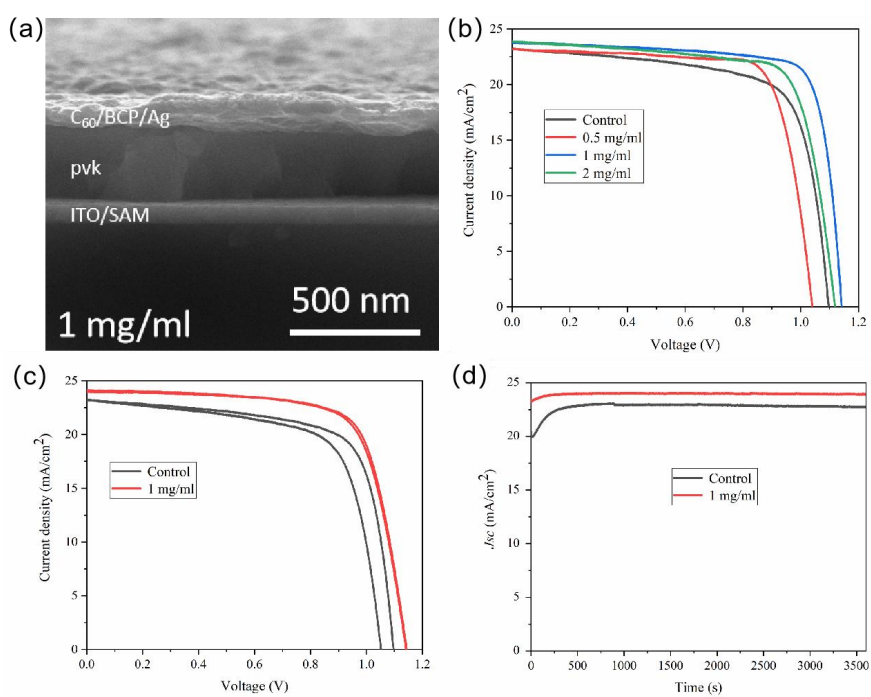

**Fig. 5.** (a) SEM cross-sectional image and (b) *J-V* curves of PSCs prepared with different concentrations of DGDMA. (c) Forward and reverse scan of PSCs. (d) Output curve of steady-state short-circuit current density at the maximum power point of different devices.

The results indicate that the PCE of the device is optimized when the concentration of DGDMA added is 1 mg/ml, achieving 21.82%, with the $V_{oc}$, $J_{sc}$ and FF all reaching their maximum values of 1.14 V, 24.11 mA/cm² and 79.30%, respectively. The statistical data on the PCE of devices with different concentrations DGDMA are shown in **Table 2**. It can be seen that compared to the control, as the concentration of DGDMA increases to 1 mg/ml, the open-circuit voltage, short-circuit current density, fill factor, and photoelectric conversion efficiency all reach their optimal levels, indicating that the addition of DGDMA can improve the morphology of the perovskite film, reduce the carrier recombination and enhance the photoelectric performance of the device. However, when the concentration of DGDMA is further increased to 2 mg/ml, the performance of the device begins to decline, suggesting that an excessively high concentration of DGDMA can generate non-radiative recombination, leading to an increase in the defect density of the device.

**Table 2.** *J-V* statistics for PSCs prepared with different concentrations of DCGDMA.

|  | $V_{oc}$ (V) | $J_{sc}$ (mA/cm²) | FF (%) | PCE (%) |
| --- | --- | --- | --- | --- |
| Control | 1.09 | 23.24 | 71.29 | 18.06 |
| 0.5 mg/ml | 1.04 | 23.22 | 76.35 | 18.44 |
| 1 mg/ml | 1.14 | 24.11 | 79.30 | 21.82 |
| 2 mg/ml | 1.11 | 23.90 | 75.01 | 19.90 |

A study on hysteresis phenomena was conducted for control and devices with 1 mg/ml DGDMA, as shown in **Fig. 5c**. In Table S3, Control-F and Control-R represent the forward scan and reverse scan of the devices for control, respectively. Similarly, 1 mg/ml-F and 1 mg/ml-R represent the forward scan and reverse scan of the devices with 1 mg/ml DGDMA, respectively. The PCE of the device modified with DGDMA was found to be 20.11% by reverse scan (20.11% by forward scan), where the short-circuit current density was 24.11 mA/cm² (24.05 mA/cm²), the open-circuit voltage was 1.14 V, and the fill factor was 73.16% (72.50%). The hysteresis phenomenon can also be observed, and it is evident that the control exhibits a more pronounced hysteresis effect. Generally, the more pronounced the hysteresis phenomenon, the more detrimental it is to the overall performance of the device.

Table 3: Statistics of J-V forward and reverse scan data for PSCs prepared with and without the addition of different concentrations of DCGDMA

|  | $V_{oc}$ (V) | $J_{sc}$ (mA/cm$^2$) | FF (%) | PCE (%) |
|---|---|---|---|---|
| Control-R | 1.05 | 23.42 | 67.78 | 16.67 |
| Control-F | 1.09 | 23.24 | 71.29 | 18.06 |
| 1mg/ml-R | 1.14 | 24.05 | 72.50 | 19.88 |
| 1mg/ml-R | 1.14 | 24.11 | 73.16 | 20.11 |

Furthermore, the study of the steady-state short-circuit current density output of the devices was conducted, as shown in Fig. 5d, the stability of the short-circuit current density over a period of 3600 seconds. It can be observed that the short-circuit current density of the prepared devices is relatively stable. However, the short-circuit current density of the control is significantly higher than that of the experimental, which is consistent with the previous *J-V*.

Finally, the distribution of photoelectric parameters of devices prepared with different concentrations of DGDMA (0 mg/ml, 0.5 mg/ml, 1 mg/ml, 2 mg/ml), each type of device includes 18 points, as shown in Fig. 6. The figure shows that compared with the control, the photoelectric parameters exhibit an overall increase with the addition of different concentrations of DGDMA, where the FF increases from 71.29% to 79.30%, an increase of 8%. When the concentration of DGDMA is added to 0.5 mg/ml and 1 mg/ml, the fluctuation of photoelectric parameters is relatively small, especially when the concentration is 1 mg/ml, the fluctuation is the smallest, indicating that the device performance is the most stable and highest. At the same time, due to the addition of DGDMA, which increases the crystallinity of the film and reduces the non-radiative recombination, the short-circuit current density also increases from 23.24 mA/cm² to 24.11 mA/cm². The open-circuit voltage also significantly increases, from 1.09 V to 1.14 V. The improvement of the three main photoelectric parameters is the main reason for the increase in device efficiency.

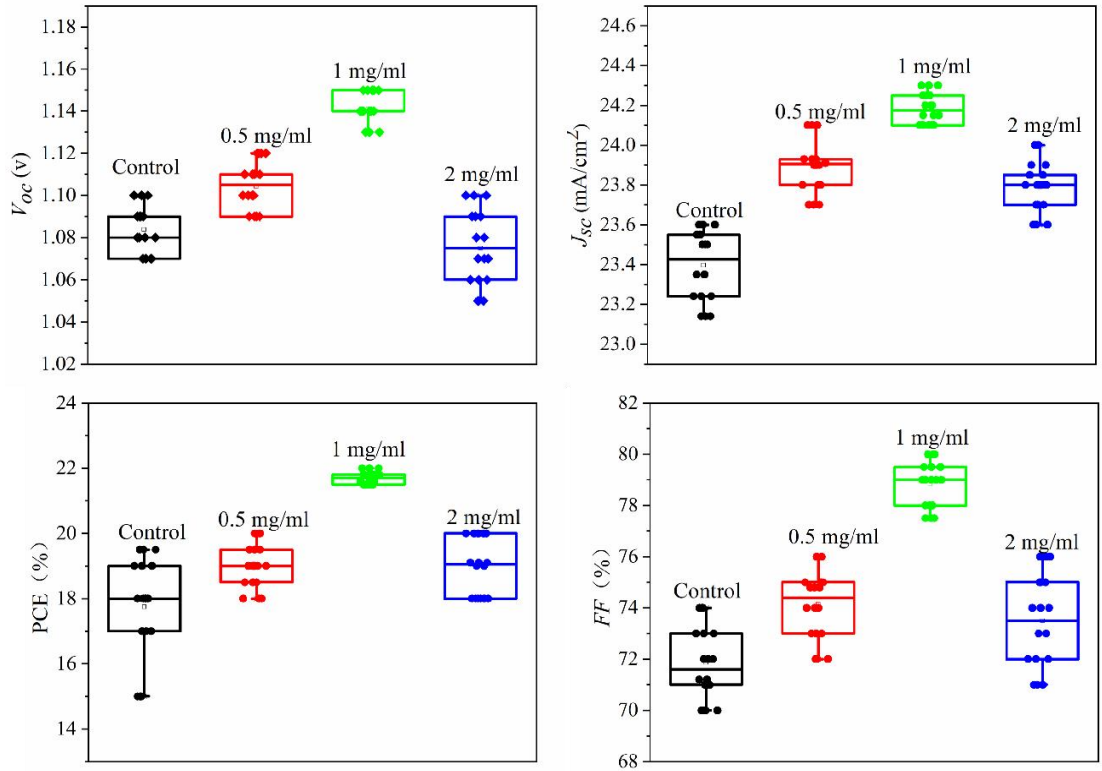

**Fig. 6.** Statistical distribution of photovoltaic parameters of PSCs prepared before and after the addition of different concentrations of DGDMA

The most important aspect to explore in perovskite solar cells is the stability. As stability improves, perovskite solar cells can be applied to people's lives. In this section, we first explore the humidity stability of perovskite films. The water contact angle was performed on perovskite films prepared with different concentrations of DGDMA (0 mg/ml, 1 mg / ml), as shown in Fig. 7a. The water contact angle of the control is 50.5°, and 62.3° for DGDMA, indicating the increased water resistance of the film after DGDMA modification.

Furthermore, an aging experiment was conducted under 60% humidity conditions for 1000 hours, and XRD were performed on the films with 1 mg/ml DGDMA and the control, as shown in **Fig. 7b**. The XRD patterns indicate that the control exhibited a strong diffraction peak at 12.8, corresponding to the (001) plane of $PbI_2$, and the perovskite peak significantly decreased, suggesting that the control underwent severe decomposition in the 60% environmental humidity, with most of it decomposing into $PbI_2$. However, only a small portion of the perovskite film with a concentration of 1 mg/ml DGDMA decomposed. In comparison with the water contact angles, this suggests that adding DGDMA can enhance the water resistance of perovskite films. The reason may be related to the typical hydrophobic alkyl groups in the DGDMA polymer, indicating a significant improvement in the humidity stability of the perovskite film.

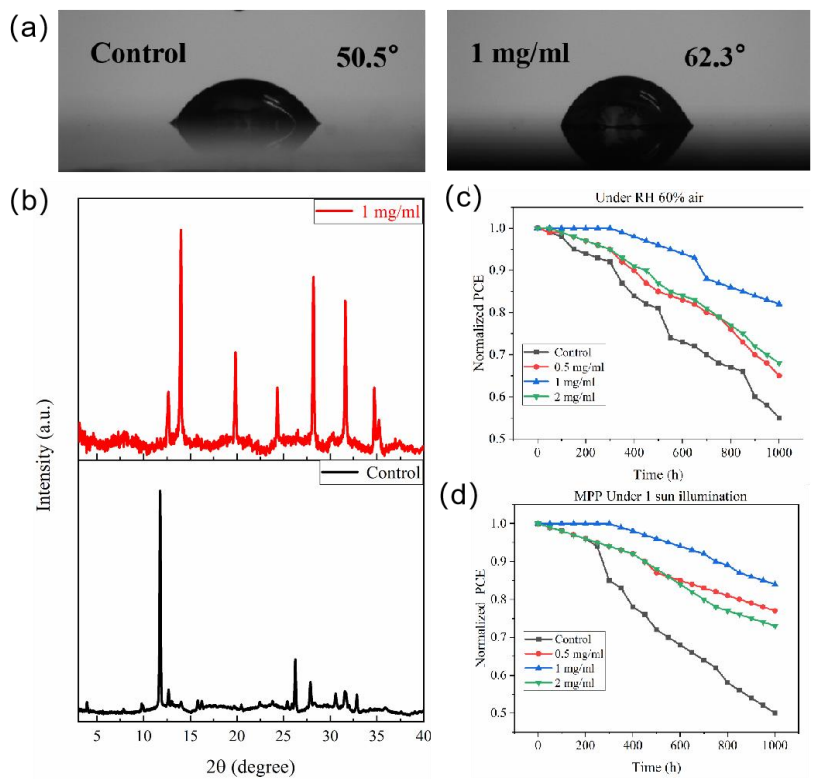

**Fig. 7.** (a) Water contact angle of PSCs prepared without and with DCGDMA. (b) XRD patterns of perovskite films. Stability of devices: (c) air humidity of 60%, (d) maximum power point for 1000 hours.

Perovskite possess soft lattice characteristics, and therefore, under the influence of harsh external environments, ions tend to diffuse, leading to a more pronounced formation of vacancies, which in turn causes the perovskite to decompose. According to reports from some researchers, the decomposition of perovskite begins at the surface of the film or at the grain boundaries, and then gradually erodes the internal crystal structure. The introduction of DGDMA can effectively inhibit the formation of vacancy defects at the surface or at the grain boundaries, thereby reducing ion diffusion and volatilization, and enhancing the stability of the film. Next, the stability of the devices is explored, mainly including humidity stability (air humidity at 60%) and light stability at the maximum power point (MPP). Stability were conducted on devices prepared with different concentrations of DGDMA (0 mg/ml, 0.5 mg/ml, 1 mg/ml, 2 mg/ml), as shown in **Fig. 7c**.

The devices only maintain 55% of the original efficiency after 1000 hours of aging under 60% relative humidity in the air. With the introduction of DGDMA, when the concentration of DCGDMA is increased to 0.5 mg/ml, the devices can maintain 66% of the original efficiency. As the concentration of DGDMA continues to increase to 1 mg/ml, the devices can maintain 82% of the original efficiency, significantly improving the humidity stability of the devices. **Fig. 7d** shows the effect of adding different concentrations of DGDMA on the optical stability of the devices at the maximum power point. After 1000 hours of continuous illumination, it can be seen that the control only maintained 50% of the original efficiency. As the concentration of DGDMA increases, when the concentration of DGDMA is 0.5 mg/ml, the devices can maintain 77% of the original efficiency. When the concentration of DGDMA is further increased to 1 mg/ml, the devices can maintain 84% of the original efficiency, indicating that the introduction of DGDMA also improves the stability of long-term continuous illumination. However, excessive introduction of DGDMA will cause the humidity stability and long-term continuous illumination stability of the devices to decrease.

## 4. Conclusion

By regulating the change of DGDMA concentration, a smooth and compact perovskite film was obtained. The results from SEM, PL and absorption spectra show that adding DGDMA can obtain the perovskite film with more flat, reduce the defect density, the non-radiative recombination and improve the device performance. From the XRD, it is found that adding DGDMA can increase the crystallinity and dactness of the perovskite film without changing the crystal structure of the perovskite material, which increases the mass of the film. The improvement of photoelectric parameters also increased the device efficiency from 18.06% to 21.82%. Finally, the stability (MPP) of the device under 60% humidity in the air and the maximum power point is tested for 1000 h. Adding 1 mg/ml DCGDMA can maintain 82% of the original efficiency under 60% humidity in the air, and 84% of the original efficiency under long time illumination. The introduction of DGDMA can greatly improve the stability of the device.

## Declaration of competing interest

The authors declared that they have no conflicts of interest to this work. We declare that we do not have any commercial or associative interest that represents a conflict of interest in connection with the work submitted.

## Acknowledgements

Project supported by the Innovation and Entrepreneurship Training Program for College Students of Xi'an University of Architecture and Technology (S202410703181, X202410703307).